\documentclass[a4paper,fleqn,usenatbib]{mnras}

\usepackage[T1]{fontenc}
\usepackage{ae,aecompl}

\usepackage{graphicx}	% Including figure files
\usepackage{amsmath}	% Advanced maths commands
\usepackage{amssymb}	% Extra maths symbols
\usepackage{pdflscape}
\newcommand{\kms} {$\mathrm{ km \; s^{-1}}\,$}
\newcommand{\msol} {M$_{\odot}$}

\newcommand{\rsol} {R$_{\odot}$}
\newcommand{\lsol} {L$_{\odot}$}
\newcommand{\about} {$\sim$}

\def\lesssim{\mathrel{\hbox{\rlap{\hbox{\lower4pt\hbox{$\sim$}}}\hbox{$<$}}}}
\def\gtrsim{\mathrel{\hbox{\rlap{\hbox{\lower4pt\hbox{$\sim$}}}\hbox{$>$}}}}

\newcommand{\degree}{$^{\circ}$}

\title[Rotation of Galactic WOs using spectropolarimetry]{Probing the rotational velocity of Galactic WO stars with spectropolarimetry}

\author[H. F. Stevance et al.]{
H. F. Stevance$^{1}$\thanks{E-mail: fstevance1@sheffield.ac.uk}, R. Ignace$^{2}$, P. A. Crowther$^{1}$, J. R. Maund$^{1}$\thanks{Royal Society Research Fellow}, B. Davies$^{3}$, G. Rate$^{1}$
\\
$^{1}$University of Sheffield, Department of Physics and Astronomy, Hounsfield Rd, Sheffield S3 7RH, UK.\\
$^{2}$Department of Physics \& Astronomy, East Tennessee State University, Johnson City, TN, 37614, USA\\
$^{3}$Astrophysics Research Institute, Liverpool John Moores University, Liverpool Science Park ic2, 146 Brownlow Hill, Liverpool, L3 5RF, UK
}

\date{Accepted XXX. Received YYY; in original form ZZZ}

\pubyear{2017}

\begin{document}
\label{firstpage}
\pagerange{\pageref{firstpage}--\pageref{lastpage}}
\maketitle

\begin{abstract}
Oxygen sequence Wolf-Rayet stars (WO) are thought to be the final evolution phase of some high mass stars, as such they may be the progenitors of type Ic SNe as well as potential progenitors of broad-lined Ic and long gamma-ray bursts. 
We present the first spectropolarimetric observations of the Galactic WO stars WR93b and WR102 obtained with FORS1 on the VLT.
We find no sign of a line effect, which could be expected if these stars were rapid rotators.
We also place constraints on the amplitude of a potentially undetected line effect. 
This allows us to derive upper limits on the possible intrinsic continuum polarisation, and find $P_{\text{cont}}<0.077$ percent and $P_{\text{cont}}<0.057$ percent for WR93b and WR102, respectively.
Furthermore, we derive upper limits on the rotation of our WO stars by considering our results in the context of the wind compression effect.
We estimate that for an edge-on case the rotational velocity of WR93b is $v_{\text{rot}}<324$ km\,s$^{-1}$ while for WR102 $v_{\text{rot}}<234$ km\,s$^{-1}$. 
These correspond to values of $v_{\text{rot}}/v_{\text{crit}}$<19 percent and <10 percent, respectively, and values of log($j$)<18.0 cm$^2$/s for WR93b and <17.6 cm$^2$/s for WR102.
The upper limits found on $v_{\text{rot}}/v_{\text{crit}}$ and log($j$) for our WO stars are therefore similar to the estimates calculated for Galactic WR stars that do show a line effect.
Therefore, although the presence of a line effect in single WR stars is indicative of fast rotation, the absence of a line effect does not rule out significant rotation, even when considering the edge-on scenario. 
\end{abstract}

\begin{keywords}
stars: Wolf-Rayet -- techniques: polarimetric -- supernovae: general  -- gamma-ray burst: general
\end{keywords}

%%%%%%%%%%%%%%%%%%%%%%%%%%%%%%%%%%%%%%%%%%%%%%%%%%

%%%%%%%%%%%%%%%%% BODY OF PAPER %%%%%%%%%%%%%%%%%%

%%%%%%%%%%%%%%%%%%%%%%%%%%%%%%%%%%%%%%%%%%%%%%%%%%%%%%%%%%%%%%%%%%%%%%%%%%%%%%%%%%%%%%%%%
%%%%%%%%%%%%%%%%%%%%% INTRO INTRO INTRO INTRO INTRO INTRO INTRO %%%%%%%%%%%%%%%%%%%%%%%%%
%%%%%%%%%%%%%%%%%%%%%%%%%%%%%%%%%%%%%%%%%%%%%%%%%%%%%%%%%%%%%%%%%%%%%%%%%%%%%%%%%%%%%%%%%
\section{Introduction}
Wolf-Rayet (WR) stars are evolved massive stars characterised by the presence of broad emission lines in their spectra.
They are the manifestation of dense winds (\about$10^{-5}$\msol$/\text{yr}^{-1}$) with velocities of several hundred to a few thousand km\,s$^{-1}$ \citep{crowther07}.
WR stars are subdivided into 3 main types: nitrogen rich (WN), carbon rich (WC) and the rarer oxygen rich (WO) stars. 
WO stars are very close to core helium exhaustion (e.g \citealt{langer12}), and are therefore the final evolutionary phase before these stars undergo core collapse, with a timescale of only a few thousand years \citep{tramper15}. 
This taxonomy also represents various degrees of stripping, with WN stars being hydrogen poor but retaining helium in their atmosphere, whereas both WC and WO are hydrogen and helium poor. 

On the whole, WR stars are excellent candidates for the progenitors of stripped-envelope core collapse supernovae (CCSNe):
The nitrogen sequence (WN) stars are anticipated to end their lives as type Ib (hydrogen poor / helium rich) SNe or IIb (transitional between type II and Ib), whereas the carbon and oxygen sequence (WC and WO) WR stars are predicted to explode as type Ic (hydrogen and helium poor) CCSNe (\citealt{filippenko97, crowther07, smartt09}). 
Among the population of type Ic SNe, a subset shows extremely broad spectral features caused by very high ejecta velocities (e.g \about 30 000km\,s$^{-1}$ in SN 1997ef and SN 2014ad -- \citealt{iwamoto00, stevance17, sahu18}) and are therefore labelled broad-lined type Ic (Ic-bl) SNe.

Additionally, some broad-lined type Ic (Ic-bl) SNe have been shown to accompany long gamma-ray burst counterparts (LGRB).
SN 1998bw \citep{patat01} was the first example of this association, while SN 2003dh \citep{stanek03} is considered the archetype of cosmological GRB-SNe. 
Additionally, LGRB SNe have been showed to prefer lower metallicity environment (e.g \citealt{modjaz08}).
One of the more likely scenarios for the production of the LGRBs is the collapsar model \citep{woosley93}, whereby a neutron star or black-hole at the centre of the exploding star accretes matter via a disk which collimates a jet that ploughs through the envelope, thus driving the explosion. 
It has also been shown by \cite{lazzati12} that Type Ic-bl SNe without GRB counterpart could also be driven by similar central engines with shorter lifetimes, which result in choked jets producing no visible GRB but still imparting the energy required for the high velocities seen in the spectra of Type Ic-bl SNe. 
This explosion mechanism was suggested for Type Ic SN 2005bf and SN 2008D \cite{maund05bf, maund08D}.
%no putting in low helium mass fraction since DEssart 2015 and Dessart 2017 say not essential to betting Ic and Ic-bl

The production of such jets through core collapse requires the progenitor star to retain a high level of angular momentum \citep{woosleyheger06}, and WO stars have been proposed as candidate progenitors for LGRB-SNe (e.g \citealt{hirschi05}).
Consequently, detecting rapid rotation in a helium deficient WR star nearing the point of core collapse would offer observational support to such theoretical models (e.g \citealt{yoon15}).
Rapid rotation in WR stars would cause wind compression effects such as described by \cite{ignace96}, resulting in aspherical winds. 
Additionally, electron scattering in WR winds is a polarising process \citep{chandra60}, and the direction of the polarisation is tangential to the last surface of scattering. 
Since the star cannot be resolved, the observed polarisation will be a sum of the components over the whole surface of the wind.
An aspherical wind, such as that of rapidly rotating stars, will then result in incomplete cancellation of the polarisation components, yielding a non-zero degree of continuum polarisation.
The presence of continuum polarisation in WR stars manifests itself as a peak or trough in $p$ across strong emission lines (called line effect -- e.g \citealt{harries98}), which is the result of a dilution of continuum polarisation by unpolarized line flux. 

Evidence for such an effect has previously been reported in Galactic, Small Magellanic Cloud (SMC) and Large Magellanic Cloud (LMC) WN and WC stars, with amplitudes ranging from \about 0.3 to \about 1 percent, and with an incidence of \about 20 percent in the Milky Way and \about 10 percent in the LMC \citep{harries98, vink07, vink17}.
The study of \cite{vink17} included spectropolarimetric observations for a binary WO star in the SMC but detected no line effect in this object. 
Because WO stars are only a few thousand years away from core collapse \citep{tramper15}, it is crucial to determine their rotational properties in order to constrain explosion models of Type Ic-bl and GRB-SNe. 

In this work we investigate the WR93b (WO3, \citealt{drew04}) and WR102 (W02, \citealt{tramper15}), whose time to explosion are estimated to be \about 9000 and 1500 years, respectively \citep{tramper15}.
They were selected on the basis that they are single Galactic WO stars accessible from the Very Large Telescope (VLT) in Chile .
The observation and data reduction processes used are described in the the following section; in Section \ref{sec:pol} we present the polarisation obtained for our targets and infer upper limits on their intrinsic continuum polarisation.
Our results and analysis and their implications for the rotation velocity and fate of WR93b and WR102 are then discussed in Section \ref{sec:disc}; finally our work is summarised in Section \ref{sec:conclusion}.

%%%%%%%%%%%%%%%%%%%%%%%%%%%%%%%%%%%%%%%%%%%%%%%%%%%%%%%%%%%%%%%%%%%%%%%%%%%%%%%%%%%%%%%%%
%%%%%%%%%%%%%%%%%%%%%%% OBSERVATIONS OBSERVATIONS OBSERVATIONS %%%%%%%%%%%%%%%%%%%%%%%%%%
%%%%%%%%%%%%%%%%%%%%%%%%%%%%%%%%%%%%%%%%%%%%%%%%%%%%%%%%%%%%%%%%%%%%%%%%%%%%%%%%%%%%%%%%%
\section{Observations and Data Reduction}
Spectropolarimetric observations of WR93b and WR102 (see Table \ref{tab:obs}) were conducted under the ESO programme ID 079.D-0094(A) (P.I: P. Crowther).
The data were collected on 2007 May 02 with the VLT of the European Southern Observatory (ESO) using the Focal Reducer and low-dispersion Spectrograph (FORS1) in the dual-beam spectropolarimeter ``PMOS" mode \citep{appenzeller98}.
The 300V grism was used in combination with a 1" slit, providing a spectral range 2748--9336\r{A}, a resolution of 12\r{A} (as measured from the CuAr arc lamp calibration). 
No order sorting filter was used.
Additionally, the observations were taken under median seeing conditions of FWHM $=0.7$". 
Linear spectropolarimetric data of our targets were obtained at 4 half-wave retarder plate angles: 0\degree, 22.5\degree, 45\degree and 67\degree.
Spectral extraction, following the prescription of \cite{maund05bf}, were performed in IRAF using routines of the FUSS\footnote{https://github.com/HeloiseS/FUSS} package as described by \cite{stevance17}.
The Stokes parameters were then calculated and the data were corrected for chromatic dependence of the zero angles and polarisation bias using FUSS. 
To improve the signal to noise ratio the data were binned to 15\r{A} which, given the breadth of the emission lines (FWHM \about 200 \r{A}), would not prevent us from resolving any potential line effect.
Intensity spectra of WR93b and WR102 were retrieved by adding the flux spectra of each ordinary and extra-ordinary ray. 
Since we did not observe a spectrophotometric standard, however, we could not calibrate the flux spectra and simply show Stokes I (in counts). 
We note that Stokes I for both our targets is presented unbinned. 
Calibrated, de-reddened X-shooter spectra of WR93b and WR102 ranging from 3000 to 25000\r{A} can be found in \cite{tramper15}.

\begin{table}
\centering
\caption{\label{tab:obs} VLT FORS1 Observations of WR93b and WR102.}
\begin{tabular}{c c c c}
\hline\hline
Object & Date & Exp. Time & Airmass \\
 & (UT) & (s)  & (mean)\\
\hline
WR93b & 2007 May 02 & 48 $\times$ 150 & 1.12\\
WR102 & 2007 May 02 & 48 $\times$ 70 & 1.06\\
BD -12\degree 5133 & 2007 May 02 & 8 $\times$ 12 & 1.14\\
\hline\hline
\end{tabular}
\end{table}

%%%%%%%%%%%%%%%%%%%%%%%%%%%%%%%%%%%%%%%%%%%%%%%%%%%%%%%%%%%%%%%%%%%%%%%%%%%%%%%%%%%%%%%%%
%%%%%%%%%%%%%%%%%%%%%%%%%%% RESULTS RESULTS RESULTS %%%%%%%%%%%%%%%%%%%%%%%%%%%%%%%%%%%%%
%%%%%%%%%%%%%%%%%%%%%%%%%%%%%%%%%%%%%%%%%%%%%%%%%%%%%%%%%%%%%%%%%%%%%%%%%%%%%%%%%%%%%%%%%

\section{Polarisation of WR93b and WR102}
\label{sec:pol}
%%%%%%%%%%%%%%%%% OBSERVED PROPERTIES %%%%%%%%%%%%%%%%%%%%%%%%%%%%%%5555
\subsection{Observational properties}
\label{sec:obs_prop}
\begin{figure*}
	\includegraphics[width=17cm]{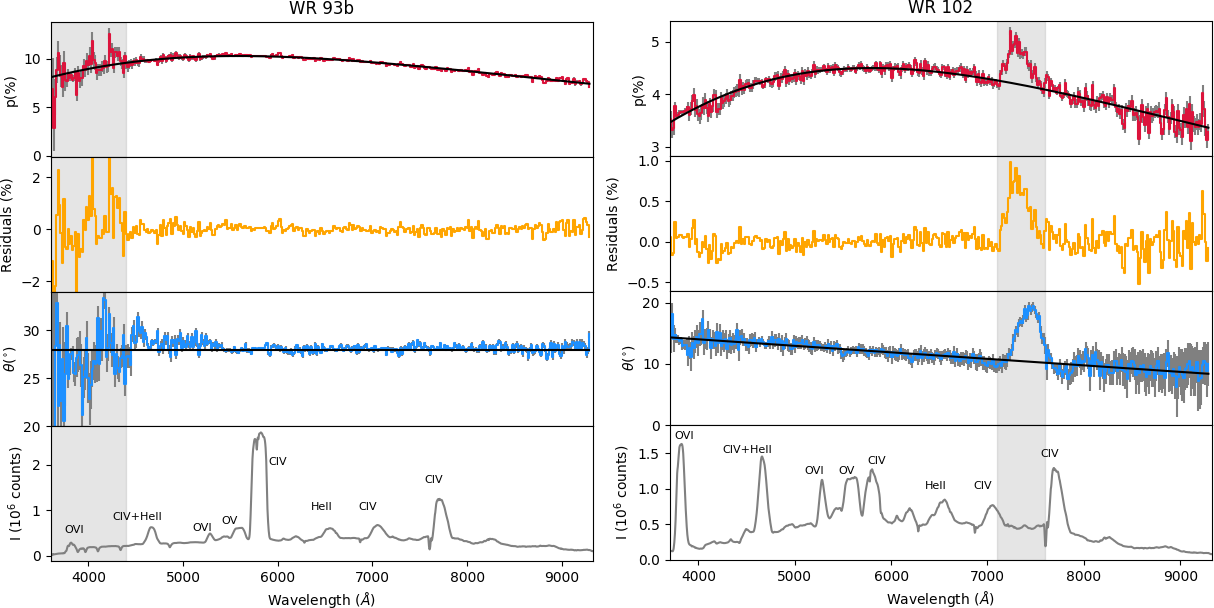}
    \caption{\label{fig:pol} Spectropolarimetric data of WR93b (left) and WR102 (right). The panels, from top to bottom, respectively contain: The degree of polarisation (red) and best Serkowski law fit (black); the residuals after Serkowski fit removal (orange); the polarisation angle (blue); Stokes I (grey) and line identification. The grey areas represent the discrepant regions of the spectrum described in Section \ref{sec:obs_prop}.}
\end{figure*}

The reduced spectropolarimetric data of WR93b and WR102 are presented in Figure \ref{fig:pol}.
%4000-4400 up to 5300
Before commenting on the characteristics of the polarisation of our targets, we highlight specific spectral regions with noisy or spurious features:
In WR93b, the degree of polarisation ($p$) and polarisation angle ($\theta$) show greater levels of noise in the blue parts of the spectrum, particularly below 4400\r{A};
in the data of WR102, a broad peak in the degree of polarisation (\about 1 percent deviation)  and polarisation angle (\about 8\degree deviation) is seen between 7100-7600\r{A}. 
The latter feature cannot be a line effect as it is located in a region of the spectrum that is devoid of strong lines, and although its origin remains unknown (the details of our investigations are given in Section \ref{sec:disc_102}), we are confident it is of spurious nature. 
Given these considerations, the analysis presented in the rest of this study was performed without the spectral region below 4400\r{A} in WR93b, and the range 7100-7600\r{A} in WR102.  

Excluding these wavelength ranges, the polarisation in the direction of WR93b and WR102 rises up to \about 4.5 percent and \about 10 percent, respectively.
Their polarisation angles have distinct behaviours: WR93b exhibits a constant polarisation angle of $\theta = 28.2$\degree, whereas the polarisation angle of WR102 shows a downward trend which can be fit with a line of the form $\theta = -1.08\times10^{-3}\lambda + 18.4$ degrees. 
A non zero $\Delta \theta/\Delta \lambda$ can be explained either by the superposition of intrinsic and interstellar polarisation \citep{coyne74} or by the presence in the line of sight of dust clouds with different particle sizes and alignments \citep{coyne66}.
The merit of these interpretations is discussed in Section \ref{sec:disc_serk}.

\subsection{7300\r{A} feature in WR102}
\label{sec:disc_102}
\begin{figure}
	\includegraphics[width=\columnwidth]{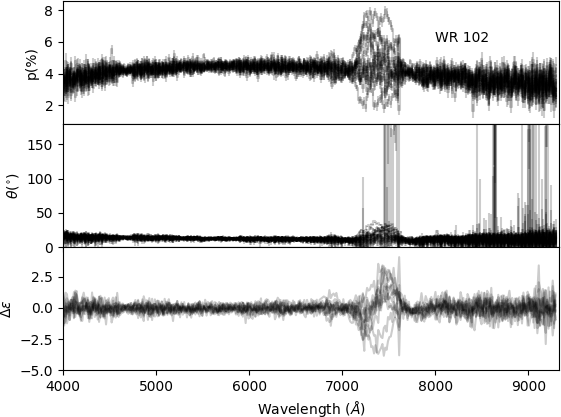}
    \caption{\label{fig:102eps} Superposed plots of the degree of polarisation (p), polarisation angle ($\theta$) and instrumental signature correction difference ($\Delta \epsilon$) as calculated for each of the 12 sets of 4 half-wave retarder plate angles of WR102.}
\end{figure}

In Section \ref{sec:obs_prop} we highlighted a prominent feature in WR102 spanning the wavelength range 7100-7600\r{A}. 
This peak was labelled as spurious and disregarded in our analysis. 
Here we detail the steps taken in investigating the nature of this feature. 

Firstly, we explored whether the spike in polarisation is intrinsic to the data. 
The fact that the feature is not associated with any strong line (as would be expected for a line effect present in the WR star) is however inconsistent with this idea. 
Additionally, plotting the polarisation calculated for each of the 12 sets of observations at 4 half-wave retarder plate angles (see Figure \ref{fig:102eps}), reveals that the shape and orientation (peak or trough) is highly variable from set to set, and no one set is consistent with another. 
To formally check whether the data is spurious we look at the difference in instrumental polarisation corrections ($\Delta \epsilon$, see \citealt{maund08}) in this region of the spectrum. 
This property is expected to be 0 for polarisation signals $\le 20$ percent, and deviations indicate that the polarisation does not reflect a real signal \citep{maund08}. 
In Figure \ref{fig:102eps} it is clearly visible that $\Delta \epsilon$ is consistent with 0 across the spectrum but deviates significantly in the wavelength range that coincides with the discrepancy. 

As the possibility that this feature was intrinsic to WR102 was ruled out, further investigation was required to understand its nature. 
We explored the possibility of a detector artefact by visually examining the 2D images, but none were found. 
Additionally, the data reduction of the polarisation standard BD -12\degree 5133 (observed immediately after WR102) showed no inconsistencies from the expected signal, and none of the WR93b data sets (observed before WR102) exhibited features such as that seen in WR102. 

Finally, we tried different approaches to the calculation of the Stokes parameters from the ordinary and extra-ordinary fluxes by:
\begin{itemize}
\item[(i)] Calculating $q$ and $u$ for each set of 4 half-wave retarder plate angles and computing the average.\footnote{Method normally used by FUSS}
\item[(ii)] Averaging the normalised flux differences for each half-wave retarder plate angle and calculating $q$ and $u$.
\item[(iii)] Averaging the ordinary and extraordinary rays for each half-wave retarder plate angle to then calculate the normalised flux differences and Stokes parameters.
\end{itemize}
As expected all these methods yielded consistent results including a visible discrepant feature around $7100-7600$ \r{A}.

Ultimately, we were not able to explain the origin of this feature as it does not seem to be either intrinsic to WR102 nor to be an issue with FORS1. 
However, since the rest of the data were unaffected we carried out our analysis disregarding the affected region of the spectrum.

%%%%%%%%%%%%%%%%%% FITS AND RESIDUALS %%%%%%%%%%%%%%%%%%%
\subsection{Interstellar Polarisation and limits on the line effect}
\label{sec:serkfits}
%In order to establish whether a line effect is present in WR93b and WR102, we must first investigate the contribution of interstellar polarisation (ISP) to the signal detected for WR93b and WR102.
%One of the first steps in spectropolarimetric studies is to quantify the interstellar polarisation (ISP) contribution. 

The dust present in our Milky Way has a tendency to align along the magnetic field of the Galaxy which causes the light that passes through these dusty regions to become polarised. 
Both WR93b and WR102 are close to the Galactic plane and are obscured by a large amount of dust, as is evidenced by the high reddening values reported by \cite{tramper15}: $E(B-V) = 1.26$ and $E(B-V) = 1.94$ mags for WR93b and WR102, respectively.
This results in high levels of ISP.

In order to retrieve the signal intrinsic to our targets it would be ideal to quantify this ISP and remove it.
To tackle this problem, one approach is to look at nearby standard stars that are intrinsically unpolarized, as any polarisation detected for these objects can then be attributed to ISP.
Unfortunately, the closest standard polarisation stars in the \cite{Heiles} catalogue were located over a degree away from our targets, making them inadequate probes of the interstellar medium between us and WR93b and WR102.

In this context a precise estimate on the ISP in the direction of our targets is not possible, however for the task at hand an absolute measure is not actually necessary.
Indeed, as we are looking for the presence of a line effect, such as that observed in \cite{harries98}, all that is needed is to investigate whether the polarisation associated with emission lines shows any deviation from the \emph{observed} continuum, which could be a blend of ISP and intrinsic continuum polarisation.
We can fit this underlying signal and remove it, to then investigate whether the residual data are consistent with noise or if a line effect can be detected.

To this end we can use our knowledge of the shape of the ISP in the Milky Way, which is described by Serkowski's law \citep{serkowski73}:
\begin{equation}
\label{eq:serk}
p(\lambda) = p_{\text{max}}\exp\bigg[-K\text{ln}^2\bigg(\frac{\lambda_{\text{max}}}{\lambda}\bigg)\bigg],
\end{equation}
where $p(\lambda)$ is the polarisation at a given wavelength $\lambda$ (in microns), $p_{\text{max}}$ is the maximum polarisation, $\lambda_{\text{max}}$ is the wavelength at maximum polarisation in microns, and K is a constant. 

We performed fits of the degree of polarisation of WR93b and WR102 with $p_{\text{max}}$, $\lambda_{\text{max}}$ and K as free parameters using a non-linear least-squares optimiser (SciPy - \citealt{scipy}). 
The best fitting parameters are summarised in Table \ref{tab:fits} and the corresponding fits are shown in Figure \ref{fig:pol}.
Should there be any intrinsic continuum polarisation, the shape of $p(\lambda)$ will remain unchanged since polarisation arising from Thomson scattering (such as can be the case in WR winds) has no wavelength dependence.
Consequently a blend of ISP and intrinsic continuum polarisation will still be described by Eq. \ref{eq:serk}, although the values of $p_{\text{max}}$ would differ from the ISP-only case. 
These fits can be subtracted to the polarisation data observed for WR93b and WR102 to yield the residuals seen in Figure \ref{fig:pol}.
No evidence of a line effect is visible in either WR93b or WR102.

\begin{table}
\centering
\caption{\label{tab:fits} Best fitting parameters of the Serkowski law fits to WR93b and WR102}
\begin{tabular}{c c c c }
\hline\hline
Object & $p_{\text{max}}$  & $\lambda_{\text{max}}$  & K\\
 & (percent) & (\r{A}) & -- \\
\hline
WR93b & 10.31 $\pm$ 0.01  &  5577 $\pm$ 18 & 1.25 $\pm$ 0.02\\
WR102 & 4.50 $\pm$ 0.01 & 5795 $\pm$ 15 & 1.30 $\pm$ 0.03\\
\hline\hline
\end{tabular}

\end{table}

In order to place limits on the maximum amplitude a line effect could have while still being undetected, we measure the standard deviation of the residual polarisation in the wavelength ranges corresponding to strong line regions, see Table \ref{tab:limits}.

\begin{table*}
\centering
\caption{\label{tab:limits} 1$\sigma$ limits on the polarisation that could remain undetected in the signal associated with the strong emission lines of WR93b and WR102. The intensity at peak relative to continuum $I_{\ell}$ as well as the derived upper limit on the intrinsic continuum polarisation are also given. }
\begin{tabular}{c c c c c }
\hline\hline
Line & Wavelength Range & 1$\sigma$ (percent)  & $I_{\ell}$ &  $>P_{\text{cont}}$ (percent) \\
\hline
\multicolumn{5}{c}{WR93b} \\
\hline
C\,{\sc iv} $\lambda 5808$ &  ($5705-5920$)   &  0.066 & 6.20 & 0.077\\
C\,{\sc iv} $\lambda 7724$  & ($7640-7840$) & 0.118 & 2.13 & 0.173\\
\hline
\multicolumn{5}{c}{WR102} \\
\hline
C\,{\sc iv} $\lambda 4659$  + C\,{\sc iv} $\lambda 4686$  + He\,{\sc ii} $\lambda 4686$ &  ($4600-4775$) & 0.05 & 2.74 & 0.068\\
O\,{\sc vi} $\lambda 5290$ &  ($5225-5360$) & 0.03 & 1.13 &  0.057\\
O\,{\sc v}  $\lambda 5590$  & ($5510-5690$) & 0.035 &  0.79 & 0.079\\
C\,{\sc iv} $\lambda 5808$   & ($5705-5930$)  & 0.05 & 1.15 & 0.093\\
He\,{\sc ii} $\lambda 6560$ + C\,{\sc iv} $\lambda 6560$ &  ($6380-6670$)  & 0.06 & 0.69 & 0.147\\
C\,{\sc iv}  $\lambda 7724$  & ($7640-7870$)  & 0.11 & 1.95 & 0.166\\

\hline\hline
\end{tabular}
\end{table*}

%%%%%%%%%%%%%%%%%%%%%% LIMITS ON POLARISATIPN %%%%%%%%%%%%%%%%%%%%%%%%%55
\subsection{Upper limit on the continuum polarisation}
\label{sec:lim_pol}
The line effect we search for is caused by a dilution of intrinsic continuum polarisation $P_{\text{cont}}$ by the unpolarized line flux. 
Under the assumption that the flux of the emission lines is completely unpolarized, the continuum polarisation is related to the amplitude of the line effect ($\Delta P$) and the peak intensity relative to the continuum ($I_{\ell}$) by the relationship given in \cite{harries98}:
\begin{equation}\label{eq:pcont}
P_{cont} = \Delta P \frac{I_{\ell}+1}{I_{\ell}}
\end{equation} 

As already stated in Section \ref{sec:serkfits}, there is no indication of a line effect in the polarisation data of WR93b and WR102, however we can make use of the limits we placed on the amplitude of a potentially undetected line effect (see Table \ref{tab:limits}) and Eq. \ref{eq:pcont} to put constrains on the continuum polarisation in our targets.
Our measurements of $I_{\ell}$ for the strong emission lines of WR93b and WR102 are given in Table \ref{tab:limits}.
An upper limit on the continuum that could remain undetected for each of these lines can then be estimated by substituting into Eq. \ref{eq:pcont} the measured $I_{\ell}$ and our standard deviation value (in the corresponding wavelength range) in place of $\Delta P$. 
The resulting values of $P_{\text{cont}}$ are given in Table \ref{tab:limits}, and represent the maximum level of polarisation that could be left unseen for each  \textit{individual line}. 
These upper limits on $P_{\text{cont}}$ vary greatly from line to line, which is not unexpected as our proxy for $\Delta P$ is the standard deviation measured for a residual signal that is dominated by noise.

%The continuum polarisation arising from Thomson scattering (such as can be the case in WR winds) has no wavelength dependence. 
In order to select a final upper limit from the values calculated we need to consider that, as previously mentioned, the continuum polarisation is not expected to be wavelength dependent.
This means that any amount of polarisation that goes undetected in one strong line will also be present in every other line and potentially be detectable. 
As a results, the effective 1$\sigma$ limits on our values of $P_{\text{cont}}$ for WR93b and WR102 will be the lowest values obtained from Eq. \ref{eq:pcont} and reported in Table \ref{tab:limits}, which do not necessarily involve the strongest emission line.
Consequently, our upper limit on the continuum polarisation $P_{\text{cont}}$ of WR93b and WR102 are 0.077 percent and 0.057 percent, respectively. 

%Therefore, although, 0.5 percent continuum polarisation could remain undetected in the C\,{\sc iv} $\lambda 7726$ line of WR102 with a 0.3 percent probability, there is about a 1 in 10 million chance that this level of polarisation would be undetected in O\,{\sc vi} $\lambda 5290$, given the respective scatter in the residuals associated with those two lines. 

\subsection{ISP and Serkowski fits}
\label{sec:disc_serk}
It is now clear that the  continuum polarisation of WR102 and WR93b, should there be any, is low ( $<0.077$ and $<0.057$ percent, respectively). 
This implies that the polarisation signal detected is overwhelmingly dominated by ISP, and our fitting parameters (summarised in Table \ref{tab:fits}) are representative of the ISP. 
Also, we can now deduce that the wavelength dependency of the polarisation angle of WR102 mentioned in Section \ref{sec:obs_prop} is likely caused by two dust clouds overlapping in our line of sight with different particle sizes and grain orientation, rather than a superposition of ISP and continuum polarisation. 

In the case of both WR93b and WR102, we find $\lambda_{\text{max}}$ to be very close to the median value of 0.545 $\mu$m observed by \cite{serkowski75} in the Milky Way.
Regarding $K$, however, quantitatively comparing our estimates to the value assumed by \citeauthor{serkowski75} is delicate since they provide no errors, and it is not clear which other values were tested and ruled out.
On the whole our values of $K$ are close to the estimate of \citeauthor{serkowski75}, and they are not statistically inconsistent with each other. 
Performing a three parameter fit of the Serkowski law rather than assume $K = 1.15$ is an approach that has been employed by multiple studies (e.g \citealt{martin99, patat15})
and is motivated by the fact that the best value of $K$ can be highly variable from target to target, even within the Milky Way (e.g \citealt{martin92}). 

Lastly, the \citeauthor{serkowski75} study also reported the relation R \about $5.5 \times \lambda_{\text{max}}$ ($\mu$m), which we can use to estimate the total to selective extinction from our estimates of $\lambda_{\text{max}}$. 
We find R = 3.1 and R = 3.2 for WR93b and WR102, respectively. 
%We find R = $3.07 \pm 0.01$ and R = $3.19\pm 0.01$ for WR93b and WR102, respectively. 
These values are consistent with the estimates of \cite{tramper15} and standard interstellar medium values.

%%%%%%%%%%%%%%%%%%%%%%%%%%%%%%%%%%%%%%%%%%%%%%%%%%%%%%%%%%%%%%%%%%%%%%%%%%%%%%%%%%%%%%%%%
%%%%%%%%%%%%%%%%%%%%%%%%%%% DISCUSSION DISCUSSION DISCUSSION  %%%%%%%%%%%%%%%%%%%%%%%%%%%
%%%%%%%%%%%%%%%%%%%%%%%%%%%%%%%%%%%%%%%%%%%%%%%%%%%%%%%%%%%%%%%%%%%%%%%%%%%%%%%%%%%%%%%%%
\section{Discussion}
\label{sec:disc}
\subsection{Rate of line effect in WR stars}

\begin{table*}
\centering
\caption{\label{tab:line_effect} Incidence of WR stars found to have a line effect as a percentage of number of WR star observed. The error are 68 percentile binomial confidence intervals. For each category N indicates the total number of stars in the sample. \textbf{References:} 1: \protect\cite{coyne88}, 2:\protect\cite{whitney89}, 3:\protect\cite{schulte94}, 4:\protect\cite{harries98}, 5: This work, 6: \protect\cite{vink07},  7: \protect\cite{vink17}  }
\begin{tabular}{l c c c c c c c c c c}
\hline\hline
Location & \multicolumn{2}{c}{All types} & \multicolumn{2}{c}{WN}  &  \multicolumn{2}{c}{WC} & \multicolumn{2}{c}{WO} & References\\
 & (\%) & N & (\%) & N & (\%) & N & (\%) & N  & \\
\hline
Milky Way & 19.4$\pm 4.8\%$  & 31 &  23.8$\pm  6.3\%$ & 21 &  12.5$\pm 8.0\%$ & 8 & 0$\pm - \%$ & 2 & 1,2,3,4,5\\
LMC & 10.3$\pm 3.3\%$ & 39 & 13.6$\pm  5.0 \%$ & 22 & 5.9$\pm  3.9 \%$ & 17 & -- & 0 & 6, 7\\
SMC & 8.3$\pm 5.4\%$ & 12 & 9.0$\pm 5.9\%$  & 11 & -- & 0 & 0$\pm - \%$ & 1 & 7 \\
\hline
Total & 13.4$\pm 2.6\%$ & 82 & 16.6$\pm 3.4\%$ & 54 & 8$\pm 3.7\%$ & 25 & 0$\pm -\%$ & 3 &\\

\end{tabular}

\end{table*}

%In Section \ref{sec:serkfits} we placed upper limits on the amplitude of the line effect that could go undetected in the noise levels of the data within the wavelength ranges corresponding to the strong emission lines of WR93b and WR102 (see Table \ref{tab:limits}).
In Section \ref{sec:serkfits} we placed upper limits on the amplitude of the line effect that could go undetected in the polarisation data of WR93b and WR102 (see Table \ref{tab:limits}).
All of these limits have an amplitude significantly smaller than the amplitudes of the line effect observed in previous studies (<0.3 percent, e.g see \citealt{harries98, vink07}), consequently it is safe to conclude that there is no line effect in either WR93b or WR102. 
Including the studies of \cite{harries98}, \cite{vink07} and \cite{vink17}, a sample of 82 WR stars have been observed with spectropolarimetry: 54 WN stars, 25 WC, stars and 3 WO stars.
This sample includes Galactic, LMC and SMC WR stars. 
Of these, 11 stars showed a line effect: 9 WN stars, 2 WC star and no WO stars.
The incidence of the line effect on WR star populations in the Milky Way, SMC and LMC is summarised in Table \ref{tab:line_effect}; the errors quoted on the percentages are the 68 percentile of the binomial confidence interval.
Across all types, 13.4$\pm2.6$ percent of WR stars exhibit a line effect, however when separated into sub-types we see that 16.6$\pm3.4$ percent of WNs showed a line effect, whereas only 8$\pm3.7$ percent of WCs and none of the WO stars did.
It is important to note, however, that these values were obtained from data sets that are not uniform in quality, and should therefore be considered with caution.

On the whole, it seems clear that younger WR stars (i.e nitrogen sequence) shows a higher rate of line effect, although the apparent statistical significance is subject to the caveat mentioned above.
This is in agreement with the study of \cite{vink11}, who pointed out the strong correlation between line-effect WR stars and the presence of ejecta nebulae.
The latter are the aftermath of a recent strong mass-loss episode undergone by the stars during a red super giant or LBV phase, and are therefore associated with WR sub-types in which hydrogen is present (i.e late WN stars). 
This is consistent with the idea that heavy mass loss also causes angular momentum loss, meaning that rapid rotation can be dissipated as a WR star further evolves. 

\subsection{Rotational velocities of WR93b and WR102}

Stellar rotation, if rapid enough, can have dramatic effects on the
surrounding winds.  \cite{bjorkman93} demonstrated
that for supersonic flow and a radial driving force, flow trajectories
from an initially spherically symmetric wind lowers the density
at polar latitudes and increases the density at equatorial ones.
This is known as the Wind Compressed Disk model.  The main results were
confirmed in hydrodynamical simulations by \cite{owocki94}. 
% However, inclusion of non-radial force components can lead to stronger polar wind flow and a depressed equatorial flow \citep{owocki96}.  

For smaller rotation speeds, \cite{ignace96} explored mild distortions
from spherical flow, called wind compression zones (WCZ).  We have
employed this model to investigate how linear polarisation depends
on the stellar rotation speed.  Details of the assumptions and approximations
adopted for this model are summarised in Appendix \ref{sec:Appendix}.

\begin{figure}
	\includegraphics[width=\columnwidth]{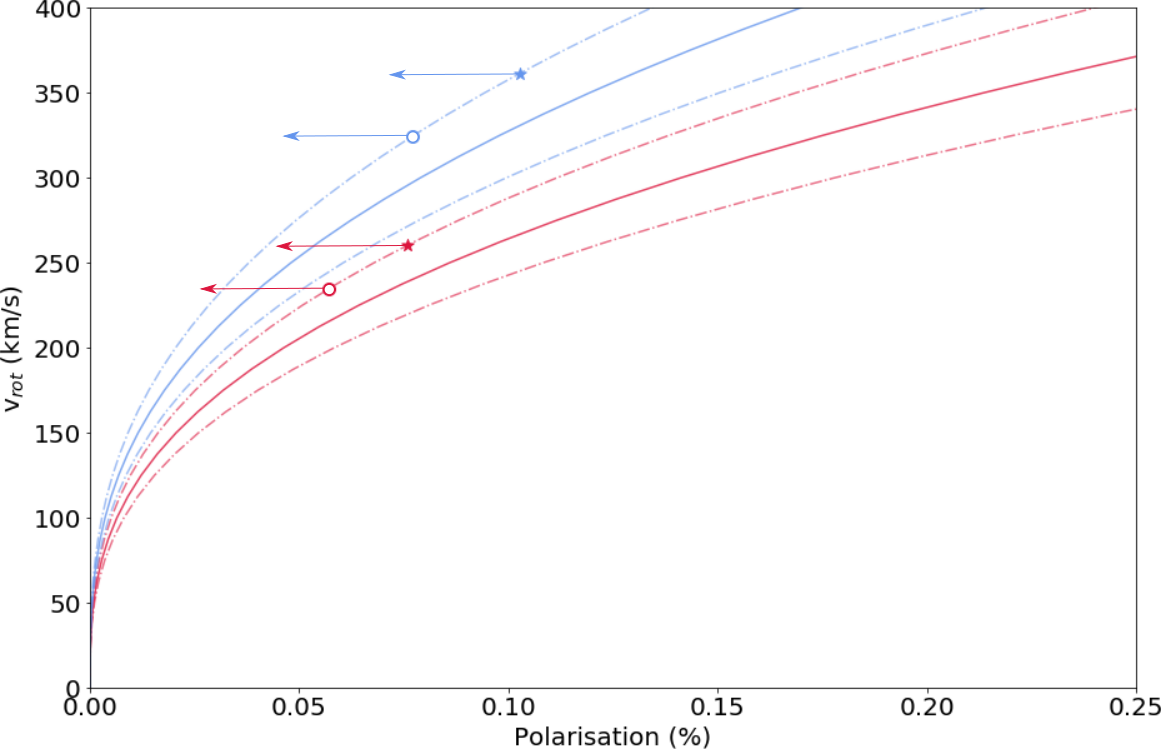}
    \caption{\label{fig:vrotpol} Relation between polarisation and rotational velocity for WR93b (blue), and WR102 (red). The dashed lines represents 1 $\sigma$ limits. The polarisation limits derived in Section \ref{sec:lim_pol} are marked by the open circles while the polarisation limits scaled by $<\sin i>^2$ are represented by the star markers.  }
\end{figure}

In order to place a limit on the rotation velocities of WR93b
and WR102, we evaluated polarisation using the WCZ approach
for optically thin scattering.  Although WR winds are optically thick
to electron scattering, the rise of optical depth % along a radial to the star
shows a sharp increase at the very inner wind owing to
the rapid drop in wind velocity, with a correspondingly fast increase
of density.  This is where multiple scattering can lead to depolarisation,
where the wind is mostly spherical, and where finite star
depolarisation effects are strongest \citep{cassinelli87}.  
We expect that most of the polarisation is formed where  the optically thin
scattering limit applies.

The WCZ models result in a prediction of the net linear polarisation
as a function of $w_{\rm rot} = v_{\rm rot}/v_\infty$.  Terminal
wind speeds of 5000 km\,s$^{-1}$ are reported for both WR93b and
WR102 \citep{tramper15}, and were used to obtain the solid lines
shown on Figure \ref{fig:vrotpol}. The errors on the relationship between degree of polarisation and rotational velocity are dominated by the errors on  $v_\infty$\footnote{ WO radii are relatively secure, despite \citet{tramper15} citing an uncertainty of $\sim$10\% in stellar temperatures. Higher (lower) temperatures will lead to smaller (larger) radii together with larger (smaller) bolometric corrections which, in turn, require higher (lower) luminosities and larger (smaller) radii for the adopted absolute visual magnitudes.} 
and are represented by the dashed lines in Figure~\ref{fig:vrotpol}.
It is worth reiterating that one underlying assumption of our model is a $\beta = 1$ velocity law.
This is a good approximation for the inner wind (see Figure 8 in \citealt{grafener05}), which is most relevant to the wind compression effect. 

Note that the curve is for a model that assumes the axisymmetric
wind is viewed edge-on.  Such an inclination maximises the observed
polarisation and minimises the limit placed on the rotation speed.
For optically thin electron scattering, the polarisation scales as
$\sin^2 i$, where $i$ is the viewing inclination angle.  Consequently,
for a given upper limit to the polarisation, the limit placed on
the rotation speed of the star is inverse to this factor.
While it is statistically unlikely to observe either WR93b or
WR102 as edge-on, it is also unlikely that either, and especially
that both, would be observed near pole-on.

In addition to considering the calculated polarisation limits (which represent the edge on case), we scaled the polarisation limits found in Section \ref{sec:lim_pol} for $<\sin^2 i> = 0.75$ to represent the expected value of a star with random inclination.  
The intersection of the polarisation limits and the upper limits on the curves of WR93b and WR102 are indicated by the markers on Figure \ref{fig:vrotpol}.  
The corresponding rotational velocities are summarised in Table \ref{tab:vrot}.

It should be noted that the wind compression effect as considered here does not include the role of additional rotational effects such as gravity darkening or oblateness. 
For the case of fast rotating B stars, \cite{cranmer95} found that the inclusion of these non-radial forces would weaken the wind compression effect, which would lead to rotational velocity underestimates of 5-10 percent.
Unfortunately no similar study exists for WR stars and such an investigation is beyond the scope of this paper, however considering the B star case our upper limits on rotational velocities may be even higher if other rotational effects were taken into account.

Lastly we point out that our upper limits on the rotational velocity of WR102 are much lower than the rotational velocity of \about 1000 km\,s$^{-1}$ inferred by \cite{sander12} from studying the shape of the emission lines in the spectrum.
The model used by \cite{sander12} used a number of assumptions and physical stellar parameters that are not consistent with the present study, which could explain the difference in values.
Most notably they assumed spherical symmetry and used physical parameters significantly different from those adopted here.
Additionally, assuming a rotational velocity of 1000\kms for WR102 would result in $v_{\rm rot}/v_{\text{crit}} = 44^{+10}_{-8}$ percent (see Section \ref{sec:vcrit})  which is much greater than observed in Galactic WR stars that did show a line effect \citep{harries98}.
These factors suggest that $v_{\rm rot}=$ 1000\kms may be an overestimate for WR102, warranting a re-evaluation of the \citeauthor{sander12} approach using  parameters based on Gaia DR2 distance.

\begin{table}
\centering
\caption{\label{tab:vrot} Summary of the upper limits on the rotational velocities of WR93b and WR102 obtained for a 90\degree inclination and a random inclination, as well as the resulting limit on $v_{\text{rot}}/v_{\text{crit}}$ and specific angular momentum $j=v_{\rm rot}R_{*}$. }
\begin{tabular}{c c c c c}
\hline\hline
 WR & Scaled p  & $v_{\text{rot}}$ & $v_{\text{rot}}/v_{\text{crit}}$ & $\log(j)$ \\
 &(\%) & (km\,s$^{-1}$) & (\%) & (cm$^2$/s) \\
\hline
\multicolumn{5}{c}{$i=90$\degree}\\
\hline
93b & <0.077 & <324 & <19 & <18.0 \\
102 & <0.057 & <234 & <10 & <17.6\\
\hline
\multicolumn{5}{c}{$<\sin i>$}\\
\hline
93b & <0.190 & <457 & <26 & <18.1 \\
102 & <0.141 & <327 & <14 & <17.7\\
\hline
\end{tabular}
\end{table}

\subsection{$v_{\text{rot}}/v_{\text{crit}}$ and specific angular momentum $j$}
\label{sec:vcrit}
In order to fairly compare the rotation of our targets to other WR stars it is useful to consider the ratio $v_{\rm rot}/v_{\text{crit}}$. 
We can calculate the value of $v_{\text{crit}}$ using 
\begin{equation}\label{eq:vcrit}
v_{\text{crit}} = \sqrt{\frac{\text{GM}_{\star}}{\text{R}_{\star}} \times (1-\Gamma_e)},
\end{equation}
where M$_{\star}$ is the stellar mass, R$_{\star}$ is the stellar radius, and $\Gamma_e$ is the Eddington factor \citep{langer98}.
Both the Eddington factor and the stellar mass are functions of the stellar luminosity \citep{schaerer92, vink15}, thus we can calculate $v_{\text{crit}}$ from the stellar luminosity (L$_{\star}$) and the stellar radius R$_{\star}$. 
We find $\Gamma_e$ = 0.12 and 0.11 for WR93b and WR102, respectively.

The recent DR2 data release of Gaia parallaxes \citep{gaia_parallaxes} allowed us to calculate new distances to WR93b and WR102: 2.3 $\pm 0.3$ kpc and 2.6 $\pm 0.2$ kpc, respectively (Rate et al. in prep). 
We can use these estimates to calculate updated values of stellar parameters from the \citeauthor{tramper15} values (see their table 4) which are summarised in Table \ref{tab:phys_param}.
The corresponding values of $v_{\text{crit}}$ are found to be: $v_{\text{crit}}=1734^{+545}_{-415}$ \kms and $v_{\text{crit}}=2286^{+528}_{-429}$ \kms  for WR93b and WR102, respectively. 
The resulting limits on the values of $v_{\text{rot}}/v_{\text{crit}}$ for a 90\degree and a random inclination are also summarised in Table \ref{tab:vrot}.
% and  calculate M$_{\star}$ using the stellar luminosity and stellar radius values given in \cite{tramper15}.
%M$_{\star}$, R$_{\star}$, L$_{\star}$ and the calculated $v_{\text{crit}}$ estimates for WR93b and WR102 are summarised in Table \ref{tab:vrot}. 
%We find $v_{\text{crit}}=1910^{+200}_{-272}$\kms and $v_{\text{crit}}=2574^{+271}_{-366}$ for WR93b and WR102, respectively.
%

\begin{table}
\centering
\caption{\label{tab:phys_param} Summary of the physical parameters of WR93b and WR102 and their calculated critical velocities.}
\begin{tabular}{c c c c c c }
\hline\hline
 WR & M$_{\star}$ & log(L$_{\star}$) &  R$_{\star}$ & $v_{\text{crit}}$ & d\\
 &\msol & \lsol  &\rsol & km\,s$^{-1}$ & kpc\\
\hline

93b & $7.1^{+2.4}_{-1.8}$ & $4.96 \pm 0.22 $ & $0.39^{+0.11}_{-0.09}$ & $1734^{+545}_{-415}$ & $2.3 \pm 0.3$\\
102 & $7.0^{+1.8}_{-1.4}$ & $4.95 \pm 0.17 $ & $0.23^{+0.05}_{-0.04}$ & $2286^{+528}_{-429}$ & $2.6 \pm 0.2$\\
\hline

\end{tabular}
\end{table}

Additionally, based on our upper limits of $v_{\rm rot}$ for our WO stars, we can calculate upper limits for their specific angular momentum ($j$), which allows us to compare WR93b and WR102 to the threshold of the collapsar scenario  ($j\geq 3\times10^{16}$ cm$^2$/s -- \citealt{macfadyen99}) as well as other Galactic WR stars \citep{grafener12}.
The values of $j$ calculated for both targets for an inclination of 90\degree and a random inclination are summarised in Table \ref{tab:vrot}. 

We find that both our limits on $v_{\text{rot}}/v_{\text{crit}}$ and $j$ are very similar to the values calculated for Galactic WR stars showing a line effect \citep{harries98,grafener12}. 
Note that the rotational velocities inferred for the WR stars in \cite{harries98} and \cite{grafener12} were calculated using spectroscopic and photometric variability, which rely on fewer assumptions than the method employed here and are therefore more robust. 
We can see that the limits placed on $j$ for our WO stars exceed the threshold for the collapsar model, and therefore cannot exclude WR93b and WR102 being LGRB progenitors.
However LGRBs are seen to prefer low metallicity environments \citep{graham13}, making this scenario highly unlikely.

Finally, these results indicate that the absence of a line effect is not necessarily synonymous of insignificant rotation. 
When investigating the presence of rotation of WR stars using spectropolarimetry, caution is therefore required when interpreting non detections as the absence of a line effect is not a direct measure of the absence of rapid rotation.

%Secondly, the incidence of $j$ values $\geq 3\times10^{16}$ cm$^2$/s  in the Galactic sample of WR stars investigated so far for the presence of a line effect is $19.4\pm4.8\%$ (not including our WO stars since our values are upper limits). 
%WR stars are the expected progenitors of type Ibc SNe, of which \about 1-9\% \citep{guetta07} are associated with LGRBs.
%Therefore the rate of Galactic WR stars satisfying the specific angular momentum condition for the collapsar model is significantly greater than the observed rate of LGRB.
%This is all the more surprising when considering that the WR in the current Milky Way are not located in the more favourable low metallicity envirnoment that most LGRBs are found in.
%This discrepancy poses the the question of the reliability of the rotational velocities derived for Galactic WR stars in previous studies, of the relevance of the comparison, and/or of the veracity of the  $j$ threshold for LGRB production. 

%%%%%%%%%%%%%%%%%%%%%%%%%%%%%%%%%%%%%%%%%%%%%%%%%%%%%%%%%%%%%%%%%%%%%%%%%%%%%%%%%%%%%%%%%
%%%%%%%%%%%%%%%%%%%%%%% CONCLUSIONS CONCLUSION CONCLUSIONS %%%%%%%%%%%%%%%%%%%%%%%%%%%%%%
%%%%%%%%%%%%%%%%%%%%%%%%%%%%%%%%%%%%%%%%%%%%%%%%%%%%%%%%%%%%%%%%%%%%%%%%%%%%%%%%%%%%%%%%%

\section{Conclusions}
\label{sec:conclusion}
In this work we presented FORS1 spectropolarimetric data of WR93b and WR102, which is the first spectropolarimetric data set obtained for Galactic WO stars.
The main results of our work are:
\begin{enumerate}
\item[1)] We find no line effect for either WR93b and WR102.
\item[2)] We deduced upper limits for continuum polarisation of $P_{\text{cont}} < 0.077$ percent and $P_{\text{cont}} < 0.057$ percent for WR93b and WR102, respectively.
\item[3)] The corresponding upper limits on the rotational velocity for an edge-on case and a velocity law $\beta=1$ are $v_{\rm rot}<324$~\kms and $v_{\rm rot}<234$~\kms, for WR93b and WR102, respectively.
\item[4)] We then found upper limits on $v_{\text{rot}}/v_{\text{crit}}$: <19 per cent and <10 percent for WR93b and WR102, respectively. 
\item[5)] Lastly we calculated upper bounds for the specific angular momentum of WR93b and WR102: log($j$)<18.0 and  log($j$)<17.6 (cm$^2$/s), respectively. 
These values do not exclude the collapsar model and we therefore cannot constrain the fate of WR93b and WR102, although the preference of LGRBs for low-metallicity environment makes this outcome highly unlikely.
\end{enumerate}

Our upper limits on $v_{\text{rot}}/v_{\text{crit}}$ and log($j$) were found to be similar to values found for Galactic WR stars showing a prominent line effect. 
Consequently this shows that the absence of a line effect is not necessarily synonymous of the absence of rapid rotation.

\section*{Acknowledgements}
The authors would like to thank the staff of the Paranal Observatory for their kind support and for the acquisition of such high quality data on the program 079.D-0094(A). 
We are grateful to Simon Goodwin and Liam Grimmett for their insight on statistics. 
HFS is supported through a PhD scholarship granted by the University of Sheffield.
RI acknowledges support by the National Science Foundation under Grant No. AST-1747658.
The research of JRM is supported through a Royal Society University Research Fellowship. 
The following packages were used for the data reduction and analysis: Matplotlib \citep{matplotlib}, Astropy \citep{astropy}, Numpy, Scipy and Pandas \citep{scipy}.

%%%%%%%%%%%%%%%%%%%%%%%%%%%%%%%%%%%%%%%%%%%%%%%%%%



\bibliographystyle{mnras}

%\begin{thebibliography}{99}
%\end{thebibliography}

%%%%%%%%%%%%%%%%%%%%%%%%%%%%%%%%%%%%%%%%%%%%%%%%%%

%%%%%%%%%%%%%%%%% APPENDICES %%%%%%%%%%%%%%%%%%%%%
\appendix
\section{Relation between rotation and polarization}
\label{sec:Appendix}

This appendix briefly reviews key ideas about Wind Compression
theory (\citealt{ bjorkman93, ignace96}), approximation formulae to characterise the density
distribution of rotating stellar winds (\citealt{ignacethesis}; Bjorkman,
private comm.), and our use of these models in relating observed
polarisation to stellar rotation speeds.

\subsection{Approximations for Wind Compression Effects}

The Wind Compressed Disk model of \cite{bjorkman93} was
developed to characterise the density and velocity distributions
for the supersonic portion of a stellar wind, under the assumption
of a radial wind driving force.  In developing observational diagnostics,
it is useful to have approximation formulae for the Wind Compression
formalism to explore parameter space more rapidly.  The following
summarises an approach appearing in \cite{ignacethesis}.

The concept of Wind Compressed Disk is based on the idea that a fluid
element of the wind, in the supersonic portion of the flow, experiences
an equatorward deflection owing to rotation.  Conserving angular momentum,
the deflection in a local orbital plane, defined by latitude of
the fluid element and centre of the star, is signified by the azimuthal
coordinate $\phi'$ \citep{bjorkman93}, with $\phi' =0$
for $r=R_\ast$.  As the fluid element moves in the wind, its
location is given by

\begin{equation}\label{eq:phi}
\phi' = \sin\theta_0 \times g(v_{\rm rot},r).
\end{equation}

\noindent where $\theta_0$ is the latitude at which the fluid element
enters the wind at $R_\ast$, and $v_{\rm rot}$ is the rotation speed
at the stellar equator.  We assume a radial wind velocity law with

\begin{equation}
w(r) = v(r)/v_\infty = w_0 + (1-w_0)\,(1-u),
\end{equation}

\noindent with $w_0 = v_0/v_\infty$, where $v_0$ is the wind speed
at $R_\ast$, and $u=R_\ast/r$.  Then

\begin{equation}
g(r)  = \Big( \frac{w_{\text{rot}}}{1 - w_0} \Big) \ln\frac{w}{w_0}
\end{equation}

\noindent with $w_{\rm rot} = v_{\rm rot}/v_\infty$.  In general,
$v_\infty$ can be a function of stellar latitude; however, this
effect is ignored here as we are mainly interested in low rotation
speed effects.  Note that $\phi'=0$ for all $r$ occurs for no
rotation and corresponds to radial wind flow.

The wind density is approximated as:

\begin{equation}
\rho \approx \rho_p + (\rho_e + \rho_p) \sin^m \theta,
\end{equation}

\noindent where $m(r)=3\tan g(r)$,  $\rho_{\rm p}$ is the density at the
pole, and $\rho_{\rm e}$ is the equatorial density:

\begin{align*}
\rho_p & = \frac{\rho_{\rm s}}{1+g(r)^2}\\
\rho_e & = \frac{\rho_{\rm s}}{\cos g(r)},
\end{align*}

\noindent where $\rho_{\rm s} = \dot{M} / 4\pi r^2 v$.
The form of $\rho_{\rm p}$ was first noted in \cite{cranmer95} (see their Appendix).

Let $K = (\rho_e/\rho_p) - 1$, then,

\begin{equation}
\frac{\rho}{\rho_p} = 1 + K(r)\sin^m\theta,
\end{equation}

\noindent which is the formulation used in calculating the wind
polarisation.

\subsection{Optically Thin Scattering Polarisation}

The calculation of electron scattering polarisation from an unresolved
source involves a volume integral over the scattering region.  
Because the wind is axisymmetric, we employ the approach of \cite{brown77},
 allowing for the finite star depolarisation effect of
\cite{cassinelli87}, and accounting for stellar
occultation.  We loosely follow the expressions of \cite{brown00}
 for the luminosity of 
polarised light, $L_Q$.  The Stokes-U component
is zero for axisymmetry, with the observer axes defined by the stellar
symmetry axis projected onto the sky.

The relevant volume integral is

\begin{equation}\label{eq:LQ}
L_Q = \int \frac{L_*}{4 \pi r^2}\, \sigma_T\, n_{\rm e}\, \frac{3}{4}\,\sin^2\chi\, \cos 2 \psi\, \Big(\frac{3K - J}{2H} \Big)\,dV,
\end{equation}

\noindent again noting that

\begin{equation}\label{LU}
L_U \equiv 0.
\end{equation}

\noindent In this expressions over volume, $\sigma_T$ is the electron
scattering cross section, $L_*$ is the stellar luminosity, $\chi$
and $\psi$ are the scattering and polarisation angles (c.f., \citealt{brown00}), and $n_{\rm e}=\rho(r,\theta)/\mu_{\rm e} m_H$ is
the wind electron density, with $\mu_{\rm e}$ the mean molecular
weight per free electron.  It is in $n_{\rm e}$ where the effects
of Wind Compression determine the source polarisation in a way
that relates to the stellar rotation.

The effect of the finite size of the star is to influence the
anisotropy of the stellar radiation field as experienced
by a scatterer in the circumstellar medium.  The parenthetical
factor appearing in the integral above involves Eddington moments
of the radiation field, and reduces to

\begin{equation}
\frac{3K-J}{2H} = \sqrt{1-u^2}
\end{equation}

\noindent accounts for the effects of the finite size of the star.

After factoring out constants, the expressions for the polarization from 
scattering can be recast as:

\begin{equation}
\frac{L_Q}{L_\ast} = \frac{\tau_0}{4\pi} \int \sin^2 \chi \,\cos 2\psi \, \frac{\text{\~{n}}(u,\mu)}{w(u)}\,  \frac{\sqrt{1-u^2}}{1+g^2(u)} \, du \,d\mu \,d\psi,
\end{equation}
with
\begin{align}
&\mu = \cos \chi,\\
&\text{\~{n}} = 1 + K(u) \sin^m \theta,\\
&\sin \theta = \sqrt{1 - \sin^2\chi \sin^2 \psi}, \\
&m  = m(u), \;{\rm and} \\
&\tau_0 = \frac{\dot{M}\,\sigma_T}{4 \pi\, R_*\, v_{\infty}\, \mu m_H}.
\end{align}

\noindent The net predicted polarisation for a rotating stellar
wind using the Wind Compression approximation formulae is
$p \approx L_Q/L_*$.

%%%%%%%%%%%%%%%%%%%%%%%%%%%%%%%%%%%%%%%%%%%%%%%%%%

% Don't change these lines
\bsp	% typesetting comment
\label{lastpage}
\end{document}